# The Digital Revolution, also for Time Scales

Claudio E. Calosso
Istituto Nazionale di Ricerca Metrologica (INRIM)
Torino, Italy
c.calosso@inrim.it

*Summary* — This work focuses on the generation of a composite clock for time scales and shows the advantages of a more recent digital approach with respect to a traditional analog one. A digital approach directly processes the information contained into the clock sinusoids, instead of the sinusoids themselves and leads to significant advantages in terms of reliability, performance, complexity, flexibility, size, power consumption and cost. A practical example based on a new digital instrument is provided to show how it is possible to combine state-of-the-art clocks from the Oscillator Imp platform at FEMTO-ST and FEMTO Engineering (Besançon, France).

I. INTRODUCTION

Many times people are so focused on physical signals that they lose sight of the information they are really interested in. This is the heritage of the past, where processing was analog and each step necessarily implied a regeneration of the physical signals. Even now, that digital electronics is well established since decades, this forma-mentis survives and represents a real limitation in taking full advantage of digital electronics and to really extend the beneficial effects of the digital revolution to time and frequency metrology, especially for what concerns clock composition and time scale generation.

By directly focusing on the information contained into the sinusoids and not on the sinusoids themselves, the digital approach allows to implement numerically the functionalities required for composing clocks, with significant advantages.

II. ANALOG VS DIGITAL APPROACH

Figure 1 shows a general scheme where several clocks at arbitrary frequencies are combined to generate a composite sinusoid, whose phase time is the weighted average of the phase times at the input.

In the analog approach (fig. 1.a), each frequency is converted into the working frequency of the micro-steppers, that keep the clocks aligned so that they can be properly combined. The alignment of the sinusoids is guaranteed by the algorithm that takes into consideration the measures of the clocks as well as additional information from primary frequency standards (in the figure a cryogenic fountain) and the data from BIPM. The algorithm sets also the weights of the combiner, that here is represented by a network of resistors. As we can see, after each step: frequency conversion, steering and composition, the sinusoids are regenerated by specific instruments leading to very complex schemes. Each instrument adds its residual noise, and performs a very specific task. Increasing functionalities can be done by adding further instruments and this represents a severe limitation. For example, we cannot find in the current practice a microstepper for each clock, because their cost-benefit is not favorable.

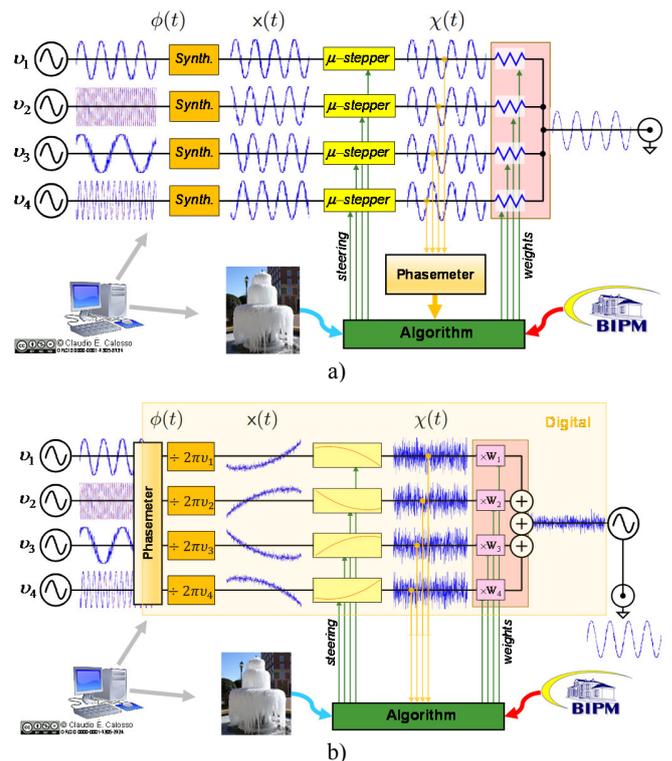

Fig. 1. Block diagrams of an analog and a digital implementation of a composite clock. In a), the sinusoids are frequency multiplied, then steered to be aligned and finally combined. In b), the equivalent functionalities are carried out numerically on the information contained into the sinusoids: phase normalization, removal of the quadratic components and then weighted average. At the end of the digital chain, and only at the end, the composed information is transposed into the phase of the output.

With the digital approach (fig. 1.b), most of the functionalities foreseen by the scheme are done numerically on the stream of data that is represented by the measures of the clocks at the input. The counterpart of the frequency conversion is represented by the phase to phase-time conversion. In practice, a frequency synthesizer is replaced by a multiplication in the code. The steering is done by subtracting from the measures their deterministic components, that, at the second order, have a parabolic trend. Here, microsteppers are replaced by parabola generators, that can be implemented in a processor with a few lines of code. Finally, a weighted average on the data streaming implements the composition of the clocks: another instrument replaced by a few lines of code.

The advantages of the digital approach are evident, once it is understood that most of the instruments and their functionalities can be replaced by a digital processor.

III. TESTS AND RESULTS

The digital approach was tested by using the Time Processor [1] with state-of-the-art oscillators and facilities at FEMTO-ST [2] and FEMTO-Engineering [3], in particular with classical RF sources (active hydrogen masers, cesium clocks and ultrastable OCXOs), microwave sources (cryogenic sapphire oscillators) and optical sources, through a frequency comb.

As a preliminary result, Fig. 2 shows the phase noise of the composition of three active hydrogen masers. There, the Time Processor measures the three masers with respect to its local oscillator. The measures are averaged and represent the error of the local oscillator with respect to the average of the three clocks. This information is fed to a controller that phase locks the local oscillator so that a physical representation of the average is obtained. The output of the instrument is measured versus a cryogenic sapphire oscillator, that acts as a reference, since its noise is negligible in the short term. The stability at one second of the composite clock is $4.5 \times 10^{-14}$, and is compatible with the short-term stability of the average of the masers.

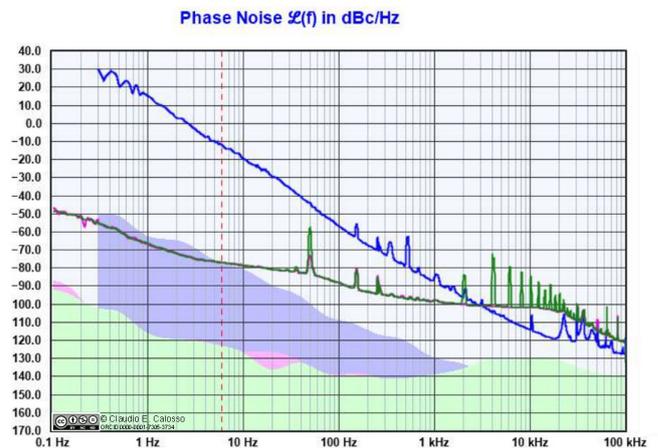

Fig. 2. Phase noise of a composite clock, where three active hydrogen masers are combined (in green). The composition is done by locking a local oscillator (in blue) to the weighted average of the masers. Carrier at 10 GHz.

IV. ACKNOWLEDGMENTS

This work is funded by INRiM internal funds, by the ANR Programme d'Investissement d'Avenir (PIA) under the FIRST-TF network (ANR-10-LABX-48-01) project, the Oscillator IMP project (ANR-11-EQPX-0033-OSC-IMP) and the EUR EIPHI Graduate School (ANR-17-EURE-00002), and by grants from the Region Bourgogne Franche Comté intended to support the PIA.